# SCUBA-2: a 10,000 pixel submillimeter camera for the James Clerk Maxwell Telescope

Wayne Holland[a*], Michael MacIntosh[a], Alasdair Fairley[a], Dennis Kelly[a], David Montgomery[a], David Gostick[a], Eli Atad-Ettedgui[a], Maureen Ellis[a], Ian Robson[a], Matthew Hollister[b], Adam Woodcraft[b], Peter Ade[c], Ian Walker[c], Kent Irwin[d], Gene Hilton[d], William Duncan[d], Carl Reintsema[d], Anthony Walton[e], William Parkes[e], Camelia Dunare[e], Michel Fich[f], Jan Kycia[f], Mark Halpern[g], Douglas Scott[g], Andy Gibb[g], Janos Molnar[g], Ed Chapin[g], Dan Bintley[h], Simon Craig[h], Tomas Chylek[h], Tim Jenness[h], Frossie Economou[h], Gary Davis[h]

[a]UK Astronomy Technology Centre, Royal Observatory, Blackford Hill, Edinburgh EH9 3HJ, UK
[b]Institute for Astronomy, Royal Observatory, Blackford Hill, Edinburgh EH9 3HJ, UK
[c]Department of Physics and Astronomy, 5 The Parade, Cardiff University, Cardiff CF24 3YB, UK
[d]National Institute of Standards and Technology, 325 Broadway, Boulder, CO 80305
[e]Scottish Microelectronics Centre, University of Edinburgh, West Mains Road, Edinburgh EH9 3JF, UK
[f]Department of Physics, University of Waterloo, Waterloo, Ontario, N2L 3G1, Canada
[g]Department of Physics and Astronomy, University of British Columbia, Vancouver, British Columbia V6T 1Z1, Canada
[h]Joint Astronomy Centre, 660 N. A'ohoku Place, Hilo, HI 96720

**ABSTRACT**

SCUBA-2 is an innovative 10,000 pixel submillimeter camera due to be delivered to the James Clerk Maxwell Telescope in late 2006. The camera is expected to revolutionize submillimeter astronomy in terms of the ability to carry out wide-field surveys to unprecedented depths addressing key questions relating to the origins of galaxies, stars and planets. This paper presents an update on the project with particular emphasis on the laboratory commissioning of the instrument. The assembly and integration will be described as well as the measured thermal performance of the instrument. A summary of the performance results will be presented from the TES bolometer arrays, which come complete with in-focal plane SQUID amplifiers and multiplexed readouts, and are cooled to 100mK by a liquid cryogen-free dilution refrigerator. Considerable emphasis has also been placed on the operating modes of the instrument and the "common-user" aspect of the user interface and data reduction pipeline. These areas will also be described in the paper.

## 1. INTRODUCTION

SCUBA-2 is a new generation, wide-field camera capable of carrying out large-scale surveys of the submillimeter sky, as well as deep, high fidelity imaging of selected regions[1]. To achieve these fundamental science goals the instrument must deliver the following:

- **Detector sensitivity dominated by the sky background.** This largely dictates the cryogenic design of the instrument in that the detectors need to operate in the 100mK regime.

- **The largest field-of-view achievable with the JCMT.** Large re-imaging optics are needed to match the 660mm diameter telescope field to a suitable array size (~100mm). This also has consequences for the size of the cryostat.

- **Detector focal planes**. The 850μm focal plane instantaneously fully-samples an area of sky, whilst the 450μm is under-sampled by a factor of 2. With over 10,000 pixels in total the volume of read-out electronics is considerably in excess of any previous instrument.

- **Dual wavelength imaging**. Imaging at two wavelengths simultaneously means that there will be two separate arrays of detectors fed via a dichroic beamsplitter.

---
[*] Email address: wsh@roe.ac.uk

In this paper we present an update on the project. Over the past 2 years the instrument design has been completed and the project is now in the verification phase of the main SCUBA-2 instrument. In parallel, prototype arrays for both operating wavelengths have been fabricated and are under test. Science-grade quality arrays are now in production for delivery of the instrument to the JCMT later this year.

## 2. INSTRUMENT ASSEMBLY AND INTEGRATION

### 2.1 Design and manufacture overview

The instrument opto-mechanical design[2,3] is driven by two principal requirements, the large field-of-view of 50 sq-arcmins, and a detector operating temperature in the 100mK regime. The large field-of-view results in extremely large mirrors (up to 1.2m across), the last 3 of which must be cooled to temperatures below 10K, in order to minimise the thermal background on the arrays. This results in a very large cryostat, the vacuum vessel of which is 2.3m high, 1.7m wide and 2.1m long, with a pumped volume of $5.3m^3$.

The cryostat itself is made up of a series of sub-systems. Immediately inside the vacuum vessel is a multi-layer insulation blanket and a radiation shield nominally operating at ~60K. This provides radiation shielding for the main optics box, which houses the cold re-imaging mirrors and will operate nominally at ~4K. Both the radiation shield and optics box are cooled by a pair of pulse tube coolers. The main optics box provides the mounting for the three cold mirrors, and the 1K box. Mounted within the 1K box are the two focal plane units that contain the cold electronics and the 450μm and 850μm arrays, which are cooled to below 100mK. The still and mixing chamber of a liquid cryogen-free dilution refrigerator cool the 1K box and arrays respectively.

To achieve the performance required the SCUBA-2 arrays must be shielded from stray light and magnetic fields. Hence the cryostat is designed with nested radiation shields and baffles, with the shields, particularly on the 1K enclosure and in the vicinity of the arrays, coated with a superconductor and high magnetic permeability material[4].

### 2.2 Main instrument assembly

The assembly and integration of SCUBA-2 is essentially split into two parts: the main instrument and the 1K enclosure (or "1K box") that houses the arrays. The main instrument consists of the vacuum vessel, 60K radiation shield and 4K optics box. The optics box houses the cold re-imaging mirrors and also provides mechanical support for the 1K box. The optics box is assembled independently from the rest of the instrument so that the mirrors can be checked prior to full system integration. Figure 1 shows the main opto-mechanical components to the 4K optics box.

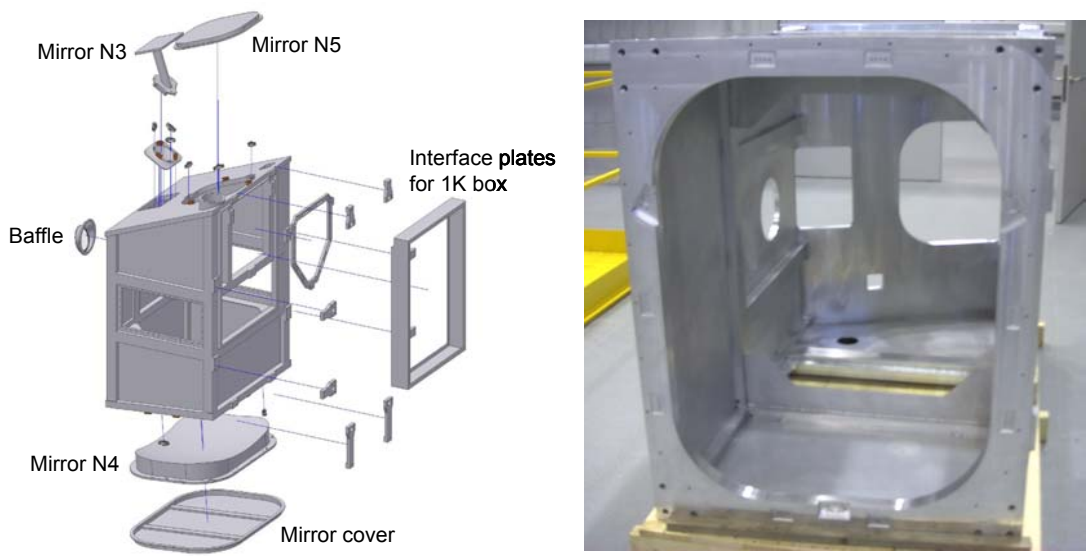

**Figure 1: (left):** The main assembly parts for the 4K optics box. These include the 3 cold mirrors, baffles and the interface plates for the 1K box and focal plane structures; **(right)** Photograph of the 4K box turned on its side (the large aperture at the front is for the N4 mirror).

Although this is the largest instrument the UK ATC has built, the design is mostly based on conventional instrument design techniques. The design is kept as modular as possible and provides good access to the important areas of the instrument. A series of photographs of the various assembly stages is shown in Figure 2.

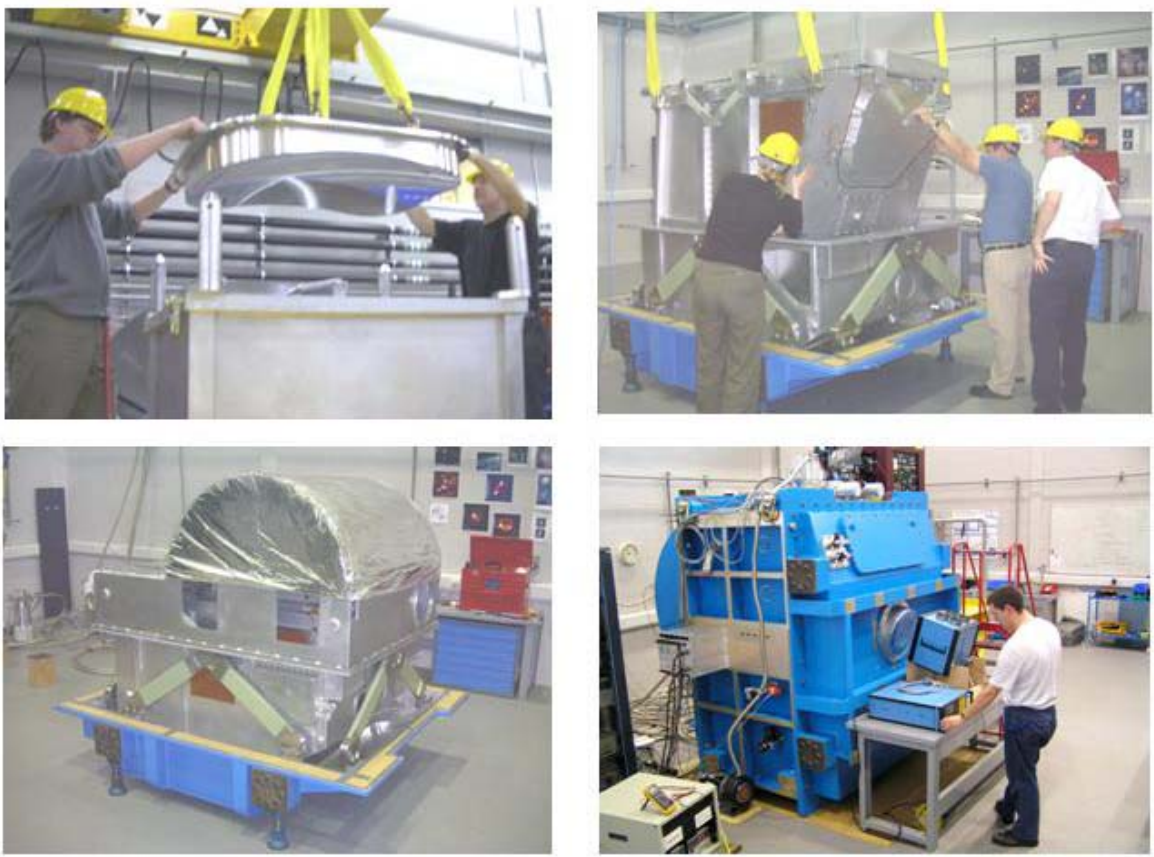

**Figure 2:** Photographs of the SCUBA-2 cryostat in various stage of assembly **(top left):** Assembling on the cold mirrors into the cryostat; **(top right):** lowering the 4K optics box into the assembled lower section of the cryostat; **(bottom left):** 60K radiation shield in place; **(bottom right):** fully assembled cryostat under test.

**2.3 1K box and focal planes**
The 1K box is an essential sub-system to create the required environment for the detector arrays. It provides (i) radiation shielding, (ii) stray light blocking (including a cold stop at the entrance), (iii) magnetic shielding for the SQUIDs, (iv) support for optical filters, (v) a dichroic to split the incoming beam onto the two focal planes, (vi) mechanical support for thermal straps and focal plane units, (vii) heat sinking for array and service wiring, (viii) support for the cold shutter (used for dark frames). The 1K box was designed and manufactured at Cardiff University and consists of an outer shell onto which 9 aluminum alloy panels are bolted. Inside the box there are four compartments primarily for stray light and magnetic field control. The box must also accurately and repeatably support and position the focal plane units with respect to the cold stop. Figure 3 (left) shows a 3-D CAD drawing of the box, highlighting the main components. There are two separate focal plane units (FPU) and each contains four array modules (see section 4.1) which are butted together to form the full field-of-view. Key elements of the FPU design include the thermal link to the DR, optical filtering and magnetic shielding. More details on the 1K box and FPU can be found in Woodcraft et al.[3] The 1K box is a separate sub-system and interfaces to the main cryostat assembly via a support frame. Figure 3 (right) shows the fully assembled 1K box installed in the main instrument. Once installation is completed the thermal links to the DR are made with the mixing chamber providing the 100mK operating temperatures for the arrays, and the still cooling the 100 kg box itself.

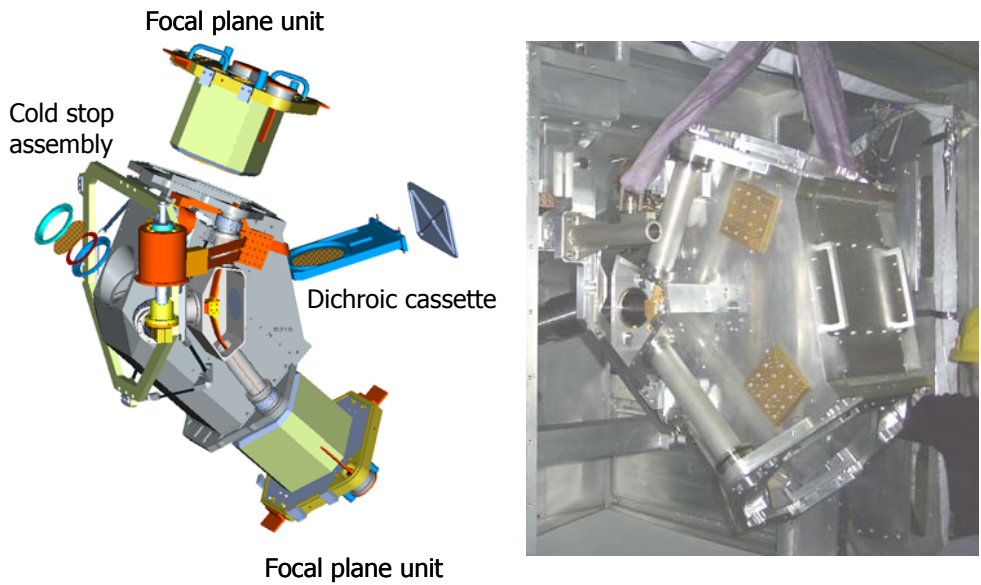

**Figure 3 (left):** 3-D CAD drawing of the 1K box showing the main components; **(right):** Fully assembled 1K box installed on support frame in the SCUBA-2 cryostat.

## 2.4 Opto-mechanical alignment and testing

As discussed in section 2.1 the cryostat is made up of a number of sub-systems. The project has adopted the policy to test as many of these sub-systems as possible separately before integration into the main instrument. Hence, for example, the vacuum vessel is tested independently at the manufacturers.

The 4K optics box includes 3 large re-imaging mirrors and the alignment of these mirrors is carried out with the instrument warm. Since the structure and mirrors themselves are all made of the same aluminum material they will thermally contract uniformly at the same rate when cooled and hence the alignment should be maintained. Fiducial marks are added to the centers of the mirrors to facilitate the use of an alignment telescope. A detailed alignment procedure has been developed but the basic method is to align the center of the optics box entrance aperture with a dummy plate presenting the cold stop (entrance to the 1K box) after reflections from the 3 mirrors. In practice the procedure was found to work well with small shifts made to mirror N4 via the use of shims.

The cold shutter is the only moving mechanism within the instrument and is critical for taking dark frames (see section 5.3.1). The shutter is mounted on the outside of the 1K box and provides a uniform background (at roughly 1K) to the arrays. The shutter motor must be shielded to prevent magnetic interference. At time of writing the shutter has undergone a limited soak test and both the mechanism and micro-switches (to determine position) have worked well.

## 3. CRYOGENIC PERFORMANCE

### 3.1 Thermal design

The overall thermal design of SCUBA-2 has been documented elsewhere[3]. In summary, the system consists of a vacuum vessel and nested radiation shields nominally at 60, 4 and 1K. There are two pre-cool tanks attached to the 60 and 4 K radiation shields. After pre-cool with liquid nitrogen the instrument is kept cold without the need for any liquid cryogens. Two Cryomech pulse tube coolers are used to cool the radiation shields and the ~300 kg of cold optics. A Leiden Cryogenics dilution refrigerator (DR) has been modified to also run with a pulse tube cooler and J-T heat exchanger, eliminating the need for a 1K pot and liquid cryogens. The still of the DR cools the 1K box, whilst the mixing chamber cools the ~30kg of focal planes to around 100 mK. Figure 4 gives a schematic representation of the cryogenic system and shows the locations of the various temperature sensors that are used to monitor performance.

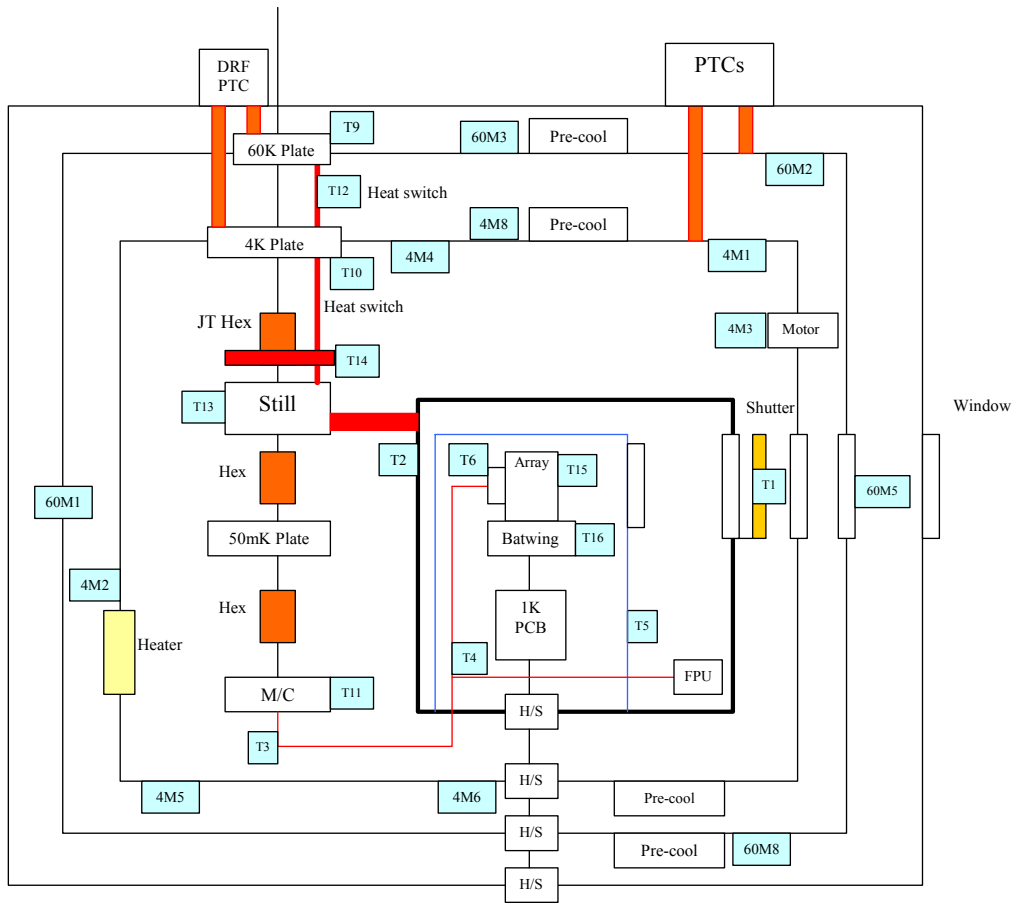

**Figure 4:** Schematic diagram of the SCUBA-2 cryogenics system showing the main thermal links and layout of thermometry (sensors are represented by filled boxes).

The DR is a key element of the system and has to cope with a substantial thermal load from the detector arrays, heat leaks down the array supports and wiring as well as radiation loading from the warmer parts of the instrument, telescope and sky. This being the case the cooling power achievable with the DR needs to be ~30μW at 65mK (with a goal of 30μW at 35mK). To cool the 1K box to <1.1 K the DR still has to have a minimum available cooling power of 5mW.

### 3.2 Thermal performance

The SCUBA-2 cryostat, complete with focal plane, 1K box, dilution refrigerator and optics box, has so far been cooled to operating temperatures on two occasions. Figure 5 shows the cooldown curve from room temperature to mK temperatures for the second laboratory run of the full system.

The key points to note are:

- Pre-cooling the system to 4K currently takes about 6 days.
- The final cooldown to mK temperatures is fairly rapid (~24 hours).
- The system has been run for 8 weeks continuously and during this time the temperatures of all stages have been extremely stable.
- There are no significant thermal gradients across the individual stages of the cryostat.

Table 1 summarises the thermal performance of the system compared to the original specification.

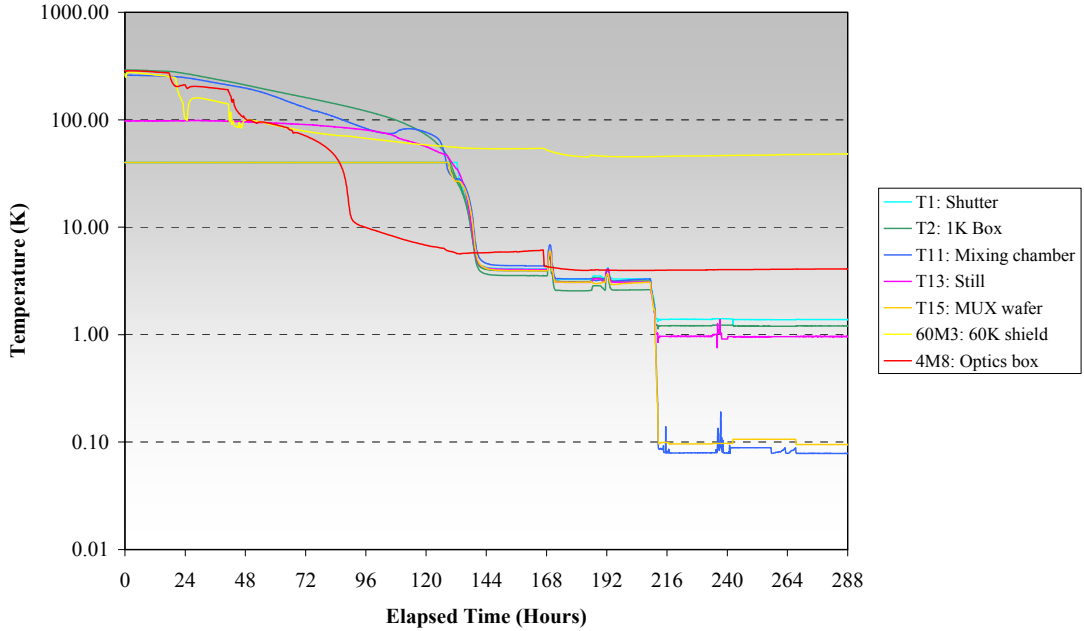

**Figure 5:** Cooldown curves for the key areas of the SCUBA-2 instrument (from March 2006). The temperature sensor labelling refers to the locations given in Figure 4.1.1.

| Parameter | Specification | Measured result |
|---|---|---|
| Radiation shield | <80 K | 50 K |
| Optics box | <10 K | 3.8 K |
| 1K box | <1.1 K | 1.3 K |
| Shutter blade | <1.5 K | 1.4 K |
| Array support | <80 mK | 80 mK |
| Pre-cool time to 4K | <4 days | 6 days |

**Table 1:** Summary of the measured thermal performance of the SCUBA-2 cryostat compared to specification.

The system meets or exceeds the specification in most areas. The 1K box is slightly too warm and it is believed that this is caused by the bolted thermal strap from the DR still not performing as well as predicted. This will be replaced by a welded strap in the near future. The high 1K box temperature also causes more of a heat leak down the sapphire isolation supports for the array. A thermal short between the 1K and mK systems limited the base temperature of the DR on the most recent run. This short has now been eliminated. The measured DR cooling power (when optimized without thermal short and leak leaks from the 1K box) is ~100μW at 40mK, significantly in excess of the specification.

There is scope for improving the pre-cool time from the present 6 days. At the moment the liquid nitrogen pre-cool runs for only ~8 hours each day. A continuous autofill system should enable the instrument to be left for long periods with minimal supervision. It will also be possible to manifold a number of $LN_2$ tanks together and run them through a single line into the instrument. Another improvement is to fill the second stage tank, then return the output of this stage into the first stage tank which is much quicker to cool. This should also be more efficient in terms of the amount of $LN_2$ used.

## 4. ARRAY PERFORMANCE
### 4.1 Array fabrication
The detailed design of the SCUBA-2 arrays as well as the fabrication has been described elsewhere[5]. The entire process of fabricating the arrays involved a number of processes. These are summarized in flowchart in Figure 6.

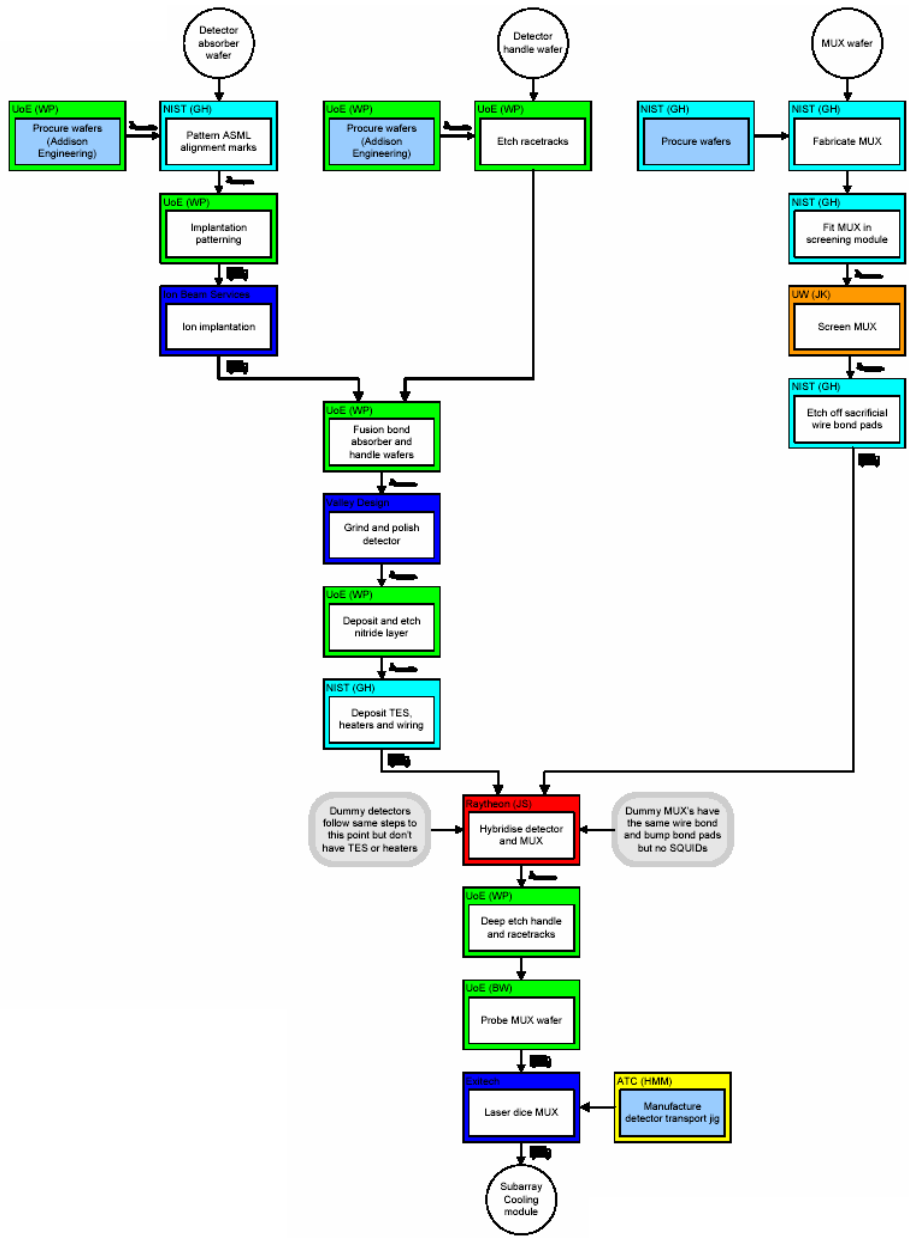

**Figure 6:** Process flow-chart for the manufacture of the SCUBA-2 sub-arrays.

As can be seen the production of a sub-array involves a large amount of processing and handling. After the above process is completed the sub-array is epoxy bonded onto a chip holder. This has been designed in the form of a "hairbrush" structure where contact is made to the underside of the MUX wafer by many beryllium-copper individual tines. This allows for differential contraction when the sub-array is cooled. It has been proven to work well with the 450 prototype array already having been successfully cooled from room temperature to 100 mK several times. Once attached to the hairbrush the sub-array is integrated into the "sub-array module" which makes electrical connection to a ceramic "batwing" PCB through aluminum wire bonds. Niobium flex cables carry the signals from the batwing to a 1K PCB which houses shielded SQUID series array amplifiers. Woven ribbon cables then take the signals to the warm electronics on the outside of the cryostat. A photograph of a sub-array module is shown in Figure 7.

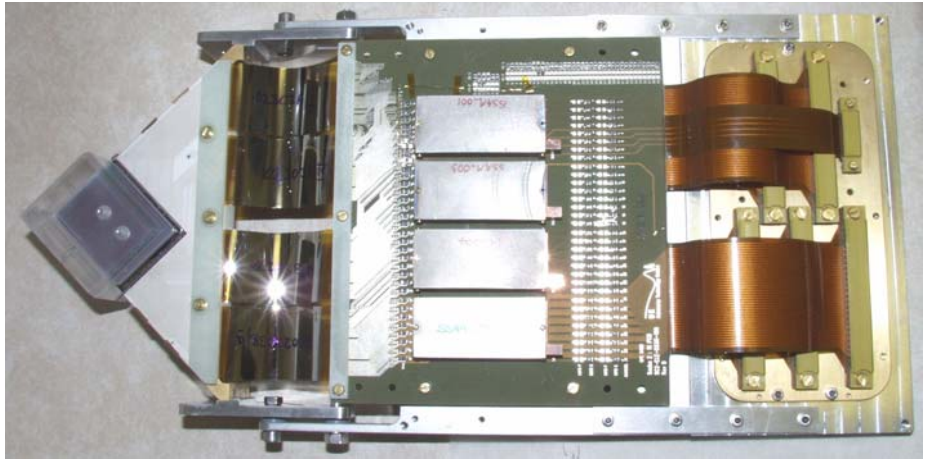

**Figure 7:** Completed sub-array module, including an array (far left with cover), "batwing" PCB, niobium flexible cables and 1K amplifier board containing SQUID series arrays.

The sub-array modules are then folded at the batwing/flex interface and assembled into a focal plane unit for test.

**4.2 Array performance**

The basic pixel design was proven by testing single pixels at NIST. This work demonstrated that the specification for the instrument could be achieved with the caveat that the next step was to prove that such performance could be scaled to an array size. Prototype sub-arrays (1280 pixels) were then developed for both 450 and 850μm. The project adopted a "plug and play" approach in which the array modules are first tested in a testbed at Cardiff before being introduced into the main system in Edinburgh.

Some of the key results so far include:

- The performance specifications for both 850 and 450 prototype arrays have been achieved.
- Magnetic shielding both in the Cardiff tesbed and main SCUBA-2 cryostat has been proven to be adequate.
- In terms of setup the system has been shown to be stable from day-to-day

Table 2 summarizes the key parameters with their measured values from both single pixel and array testing. The results refer to measurements made in the Cardiff testbed. The prototype array testing programme is more fully described by Woodcraft et al.[6]

| Parameter | 450μm | | | 850μm | | |
|---|---|---|---|---|---|---|
| | Specification | Single pixel | Proto array testing | Specification | Single pixel | Proto array testing |
| $T_c$ (mK) | 150 – 170 | 193 | 175 | 120 – 140 | 133 | 130 |
| Total power (pW) | 200 – 230 | 267 | 460 | 40 – 60 | 57 | 60 |
| G (nW/K) | ~5 | 5.2 | 9.0 | ~1.5 | 1.6 | 2.0 |
| Optical NEP (W/√Hz) | <29 × 10$^{-17}$ | 9.7 × 10$^{-17}$ | 14 × 10$^{-17}$ | <7 × 10$^{-17}$ | 3.5 × 10$^{-17}$ | 2.5 × 10$^{-17}$ |
| $\tau_e$ (msec) | <1.5 | 0.2 | 0.6 | <2.8 | 0.6 | 1.0 |

**Table 2:** Summary of the measured array parameters compared to tests on single pixels and to the original specification. The NEP values on single pixels were determined electrically from I-V curves.

Work is now underway to evaluate the prototype array performance in the main SCUBA-2 cryostat. Figure 8 shows an example set of detector I-V ("load") curves as a function of the pixel heater setting and the corresponding power as a function of voltage plot for a selected pixel. The load curves show the characteristic shape for a transition edge sensor. The power plot clearly shows that when a TES device is operated in the transition the electrical power is constant and does not vary with detector bias.

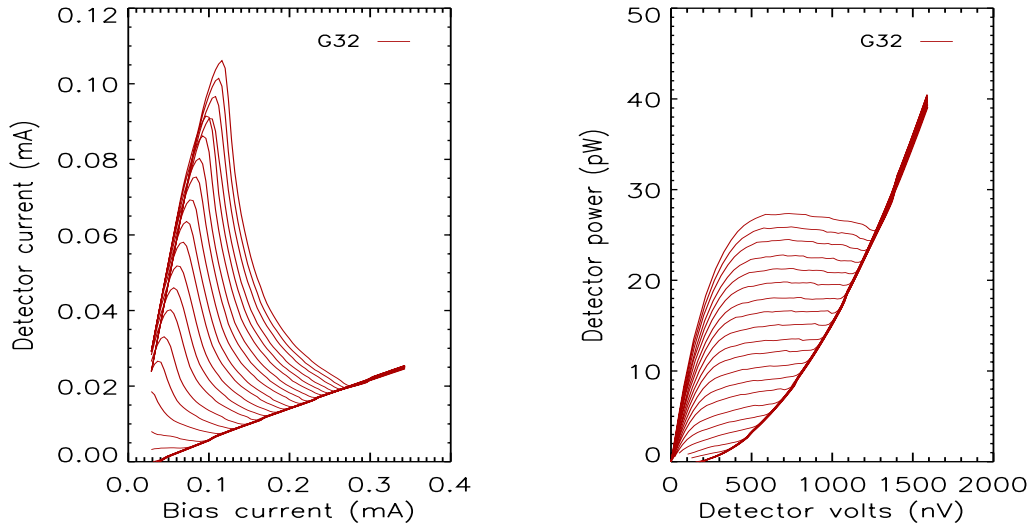

**Figure 8 (left):** An example set of detector I-V ("load") curves as a function of the pixel heater setting; **(right):** The corresponding electrical power as a function of voltage plot for a selected pixel

The testing programme so far has shown the specification can be achieved on an array scale. The project is now developing science-grade arrays to fully populate the two instrument focal planes.

## 5. TELESCOPE INFRASTRUCTURE, COMMISSIONING AND OPERATIONS

SCUBA-2 is scheduled to be delivered to JCMT in November 2006. At this point the instrument will have one science-grade array at each wavelength (i.e. each focal plane will be a quarter populated).

### 5.1 Telescope infrastructure

The size of the cryostat means that SCUBA-2 cannot be easily accommodated on the telescope. In addition, it is necessary to re-image the large telescope field (some 650 mm in diameter at the Cassegrain focus) to a size commensurate with the SCUBA-2 focal planes (around 120mm). These constraints result in a location for the instrument on the telescope that is awkward (see Figure 9 left). Hence, to accommodate SCUBA-2 and supporting equipment the telescope is undergoing considerable modification. The "SCUBA-2 Infrastructure Project" was developed to provide these facilities. As part of the implementation of the project a telescope shutdown is required lasting six months (Feb – Aug 2006). Key elements of the project are:

- A new gallery above the control room to site the support equipment for SCUBA-2 including closed cycle cooling equipment, dilution refrigerator rack and all electronics other than the MCE array controllers. This gallery was installed prior to the shutdown and was ready for the installation of power supplies and coolers as part of the shutdown work (see Figure 9 right).

- The receiver cabin will house a new Tertiary Mirror Unit and a set of relay optics to reduce the SCUBA-2 input beam to a size that can pass unvignetted through the left elevation bearing. To house these 'cabin mirrors' significant redevelopment of the receiver cabin is required. This includes relocation of existing equipment and modification of the cabin bay next to the bearing.

- The left hand Nasmyth platform that serviced the original SCUBA and allowed access to the receiver cabin has to be virtually eliminated to mount the re-imaging optics and the instrument itself. Key work during the shutdown is removal of the existing platform and rebuilding of the receiver cabin access walkway. A further part of this work is extending the mezzanine platform below to allow access to SCUBA-2 when installed.

- SCUBA-2 requires a mounting frame to support the instrument and the re-imaging "Nasmyth mirrors". This structure is being design and fabricated by contractors and will be installed by JAC staff towards the end of the shutdown.

- The mass of SCUBA-2 (~ 4 tons) is too heavy to be supported on the antenna and carousel floor by conventional means. As a result a significant part of the project is to provide dedicated handling equipment to install and remove the instrument. The floor requires strengthening and the installation of a rail system for a handling cart. The reinforcement is underway with the handling cart arriving towards the end of the period.

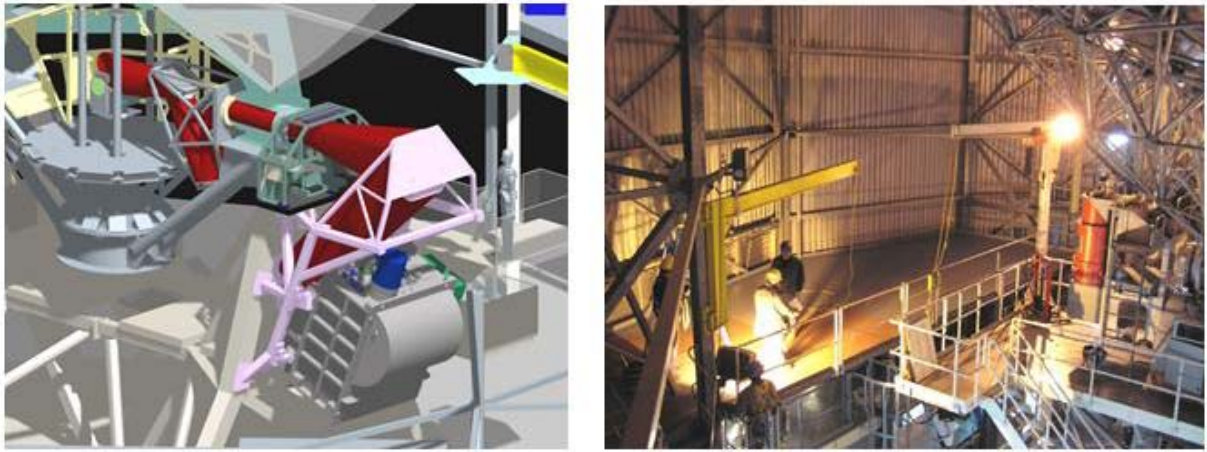

**Figure 9 (left):** 3-D CAD model of the optic relay from the tertiary mirror to the cryostat location; **(right):** The new gallery above the control room to site the support equipment for SCUBA-2 including closed cycle cooling equipment, dilution refrigerator rack and all electronics other than the MCE array controllers

### 5.2 Commissioning

The installation and commissioning of SCUBA-2 will be divided up into a number of phases as summarized in Table 3. A considerable amount of preparation work will be carried out at the telescope prior to the arrival of the instrument (see section 5.1). In addition to the telescope infrastructure work the SCUBA-2 warm mirrors will have also been assembled and aligned prior to the instrument arrival (final alignment checks will be undertaken with the instrument in place). The instrument will be delivered with (most likely) one sub-array at each wavelength and will be "upgraded" to full array complement once all sub-arrays have been tested. Hence an upgrade period has been included in this plan.

| Phase | Description and main goals | Estimated Duration |
|---|---|---|
| Hilo arrival and re-orientation | Check shipment is in tact, re-orientate cryostat as necessary and re-assemble backshells; transport to summit | 2 weeks |
| Summit assembly and integration | Assemble and install cryostat and ancillary equipment | 3 weeks |
| Cooldown and functional check | Pump out, cooldown and perform basic functionality test of arrays (in dark), check mechanisms and data acquisition, telescope actions etc | 4 weeks |
| On-sky commissioning | Demonstrate basic functionality of all observing modes | 4 weeks |

| Astronomical image quality and calibration | Image known sources to demonstrate imaging capabilities, astrometry, noise, calibration etc | 3 weeks |
|---|---|---|
| Upgrade | Upgrade focal planes to full array complement. Time includes warm-up and cooldown. | 4 weeks |
| Re-commission and Final release | Re-commission instrument using all sub-arrays and carry out acceptance and release to community. | 2 weeks |

**Table 3:** Summary of the main phases for the commissioning period for the instrument. It is anticipated that the total commissioning time, including array upgrade, will take about 5 – 6 months.

### 5.3 Operating modes
SCUBA-2 will provide JCMT with the following observing modes:

- **Imaging mode**: observing regions of sky equivalent to the array field-of-view or mosaicing together offset frames to produce an image up to a few arrays in size
- **Survey (or scan) mode**: mapping large areas of sky, potentially up to tens of degrees at a time
- **Spectroscopic/polarimetric mode**: Imaging polarimetry and medium resolution spectroscopy will be possible with additional hardware

The signal output is DC-coupled to the signal-processing electronics leading to significant efficiency improvements, since half of the integration cycle is not spent on blank sky. There is hence no need for sky chopping resulting in better image fidelity and sensitivity to source structure on ALL scales. In principle, it also gives a lower confusion limit i.e. by not continuously subtracting two images of the sky.

#### 5.3.1 Imaging modes
STARE is a "point-and-shoot" mode, in which SCUBA-2 will "stare" at an area of sky equivalent to the field-of-view for a specified period of time. Multiple fields can be mosaiced together to form a larger image as needed. To be effective for carrying out *deep imaging* (e.g. down to the confusion limit) the RMS noise in the image should integrate down with the square root of the integration time. As mentioned above SCUBA-2 will be a DC-coupled system and hence any excess ("1/f") noise from the detectors or electronics may cause a limit as to how deep an image can be obtained. In an effort to compensate for this 1/f noise a cold shutter will take "dark frames" but there may still be a residual noise floor for deep observations.

In addition to 1/f noise, the brightness of the atmosphere compared to the astronomical source is such that STARE also requires a highly accurate flat-field. The pixels will all have slightly different sensitivities and to ensure the astronomical images reflect real source structure these pixel-to-pixel variations have to be calibrated out. The accuracy required for the flat-field depends on observing mode and integration time but is most severe for STARE (estimated to be 1 part in $10^7$ for a 1-hour observation). There are also potentially several factors that can cause the flat-field to vary. These include drifts in the electronics (must be made as common mode as possible), 1/f noise in the detectors or SQUIDs, variations in the detector responses as a function of background power (i.e. caused by relative changes in the bias setting). Hence, any flat-field changes will be monitored and corrected in real time (e.g. using the cold shutter). However, the accuracy required may still not be achievable in practice for ultra-deep (or even reasonably shallow) observations.

For the case that interleaved dark and STARE frames cannot be used to make deep images an alternative technique called DREAM will be used. DREAM is the Dutch REal-time Acquisition Mode[7] and the technique makes use of the JCMT secondary mirror which can perform rapid motion in two dimensions. The mirror sweeps out a suitable pattern while the instrument takes rapid data frames. The result is that each bolometer records a small map of the sky overlapping in area with its neighbouring bolometers as shown in Figure 10. A least-squares technique is then used to solve for the combined map and the relative bolometer zero points. Both STARE and DREAM mode are being developed for SCUBA-2. The optimum mode for a given type of observation will be determined during commissioning at the telescope.

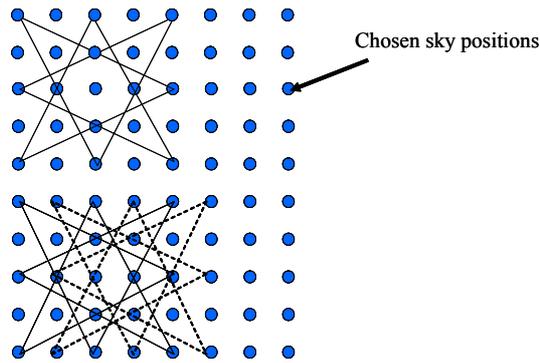

**Figure 10:** Illustration of the DREAM observing strategy. Secondary mirror is driven so that each bolometer makes a mini-map. The mini-map of each bolometer overlaps with its neighbours.

**5.3.2 Survey (scan) mode**

One of the major SCUBA-2 goals is to conduct unprecedented wide-field surveys of the sky. To map areas significantly larger than the footprint of the SCUBA-2 arrays it will be most efficient to scan the telescope. In order to be able to recover large-scale structures in the presence of slowly-varying baselines (caused primarily by sky emission, extinction, and instrumental 1/f noise) the scan pattern must modulate the sky both spatially and temporally in as many different ways as possible. Spatial modulation is achieved by scanning the same region at a number of different position angles, or cross-linking. Temporal modulation is incorporated by visiting the same region on different time-scales. For most applications it would be desirable for the effective integration times across the mapped region to be as flat as possible. Finally, it should be easy to define the shape of the mapped region. A number of scan patterns have been considered ranging from simple rectangular raster maps, to more complicated Lissajous curves. A pattern called PONG map has been chosen as the basic scan mode and this will be provided at first light for the instrument.

A PONG map fills in a rectangular region using straight constant velocity scans. An inclined scan angle is chosen with respect to the native axes of the rectangle, and once a border of the rectangle is encountered the scan direction is reflected. A simple square subset of these patterns may be parameterized which provides orthogonal cross-linking by three variables: (i) the grid count, or odd number of scan lines in one dimension of a square grid; (ii) the grid interval, or the space between each grid line; and (iii) the grid angle which gives the rotation of the PONG pattern with respect to the telescope azimuth. In a single pass through this pattern the time between the cross-linking scans occur at different times depending on the position in the map. If the grid interval is chosen to be a fraction of the array size the coverage of the PONG map is uniform. Figure 11 shows an example of a PONG scan with grid count = 7 and grid interval = 4 arcmin. If the telescope were slewed through this pattern at a velocity of 200 arcsec/sec it would take approximately 45 seconds.

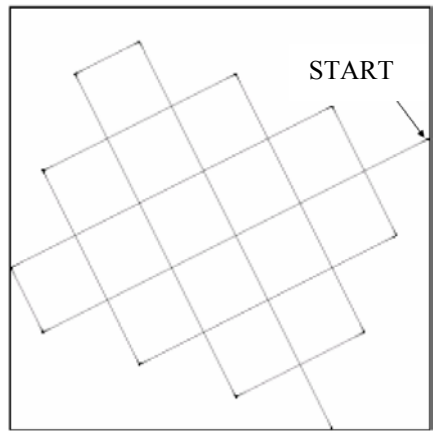

**Figure 11:** An example of a PONG scan with grid count = 7 and grid interval = 4 arcmin. If the telescope were slewed through this pattern at a velocity of 200 arcsec/sec it would take approximately 45 seconds.

## 5.4 Data Flow and Pipeline Infrastructure

The 200-Hz sampling rate for SCUBA-2 translates to a data rate of approximately 2 MB/s at each wavelength or 200 GB for a 12-hour observing night. Each sub-array will have its own dedicated data acquisition (DA) computer which will write the raw frames to a local disk. The data processing pipeline will use the established ORAC-DR pipeline infrastructure. ORAC-DR is a proven pipeline technology employed on the JCMT, UKIRT and the AAT and will be able to handle data from SCUBA-2[8,9].

The data reduction pipeline will produce accurately-calibrated and scientifically-meaningful images, free from imaging artifacts (e.g. cosmic rays, bolometer drifts). There will be four pipelines running simultaneously at the telescope, 2 for each wavelength (see Figure 12 for an example of the data flow for one wavelength). One of these is the Quick-Look (QL) pipeline, designed to provide rapid feedback to observers, and the other is the data-reduction (DR) pipeline designed to produce scientifically-meaningful (i.e. calibrated, astrometrically correct) images. The QL pipeline is similar to the DR pipeline, differing only in the recipes applied to the incoming data. Simple de-spiking, co-adding and re-gridding onto a preferred coordinate frame (e.g. RA/Dec) will enable the observer to make online assessments of the data.

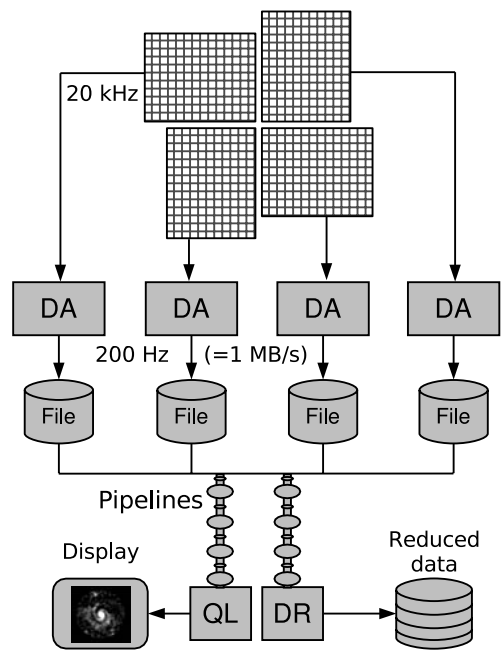

**Figure 12:** Schematic view of the data flow from one of the SCUBA-2 bolometer arrays showing the two pipelines running concurrently. The data from each sub-array is handled by a single DA computer, which the pipelines read from to produce images.

The pipelines will run on separate computers, reading the raw data from the remote disks associated with the DA computers and processing them locally to minimize network-induced lag. The pipeline will produce high quality images in so-called "near real time". The definition of near real-time is that it will take no longer to process the data than it took to obtain the data. For the QL pipeline, this means that the data must be processed and displayed before the next observation is complete. Testing of the QL pipeline shows that current hardware is capable of keeping up with the data rate. Images generated in DREAM/Stare mode will be processed and displayed as they come in, on timescales of order 1 second. Scan-map data taken at 200 Hz will be processed and displayed before the next scan is completed. For the DR pipeline data from a single 12 hour night must be processed by the start of the next observing night.

The pipeline will have an offline mode which will make use of calibration data derived from the whole night (rather than using the most recent) as well as allowing for some flexibility in control by users. The offline mode will also allow data from multiple nights to be combined[10] and users will have full access to the 200 Hz data in all modes. The pipeline will also monitor the incoming data to provide diagnostic information for assessing instrument performance, observing conditions and data quality. The QL pipeline will provide direct visual feedback when bright sources are observed, while other measures of performance will also be calculated and made available to a stripchart plotter. The stripchart tool is configurable to monitor various parameters and pass updates to one or more plots for tracking variations. The high sensitivity of SCUBA-2 and the production of fully-sampled images on timescales of 1 second will allow for significant improvements in the calibration accuracy of the pipeline. For example, bright point sources can be used to provide updates to the telescope pointing, estimates of the submillimetre seeing and checks on the telescope focus.

**5.5 Legacy survey programmes**
One of the key scientific drivers for SCUBA-2 is to carry out large-scale surveys of the submillimeter sky. Over the past year or so a number of "legacy-style" survey programmes have been developed. Seven of these surveys have now been approved. In summary these surveys are:

- Galactic Plane survey: 200+ sq-degs to a depth of 4mJy at 850μm in 330 hrs
- Gould Belt survey: Mapping of G-B molecular clouds (~500 sq-degs) to 10mJy (850) in 320 hrs (inc. POL)
- Debris Disk survey: Survey of 500 stars to 0.5mJy (850) in 370 hrs
- Local Galaxy survey: Imaging of 150 galaxies down to 1mJy (850) in 75 hrs
- Cosmology survey (850μm): 20 sq-degs (several fields) to 0.7mJy in 630 hrs
- Cosmology survey (450μm): 0.7 sq-degs to 0.5mJy in 490 hrs
- "All sky survey": 4500 sq-degs down to ~30mJy at 850μm

Figure 13 summaries the area and depth for each of the surveys. The "All-sky" survey is a programme to utilize less than perfect observing conditions and will be carried out as a poor weather backup project. These survey programmes are due to start around mid-2007 and will take up about 50% of the available telescope time over at least a period of 2 years. Open-time applications and engineering time will consume the rest of the time.

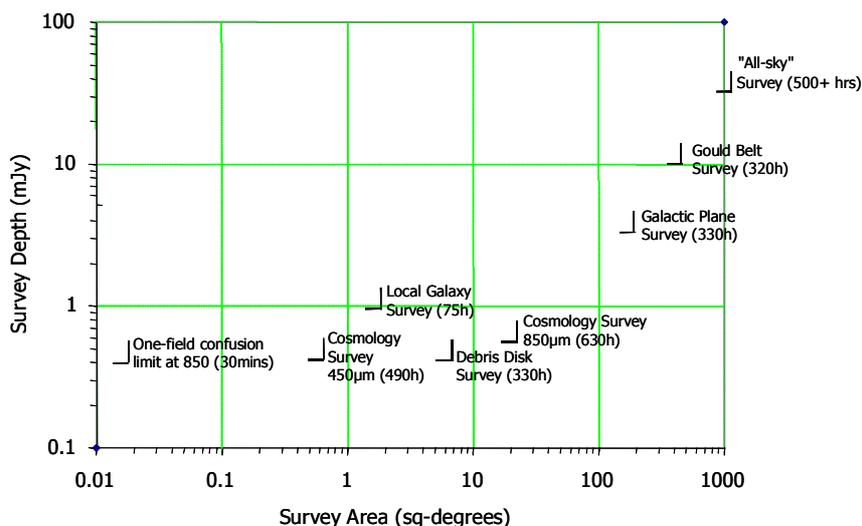

**Figure 13:** Summary of the approved SCUBA-2 Legacy programmes for the JCMT in terms of total area to be covered to a certain depth.

## 6. SUMMARY

SCUBA-2 is currently in the instrument evaluation phase prior to delivery to Hawaii in late 2006. The 10,000 pixel camera represents a huge leap in imaging capability for the submillimeter and will enable (particularly) survey-type programmes to be undertaken for the first time in this wavelength regime.

SCUBA-2 represents a major innovation from current submillimeter instruments. The science applications for such an instrument are tremendously exciting and very broad-based, ranging from the study of Solar System objects to probing galaxy formation in the early Universe. *SCUBA-2 will map large-areas of sky up to 1000 times faster than the current SCUBA*. The improved sensitivity and imaging power will allow the JCMT to really exploit periods of excellent weather on Mauna Kea. Undertaking wide-field surveys with SCUBA-2 are vital to fully exploit the capabilities of the new generation submillimeter interferometers. Finally, the new technology has applications beyond SCUBA-2, and thus represents a major strategic investment on behalf of the JCMT and instrument funding agencies.

## ACKNOWLEDGEMENTS

The SCUBA-2 project is funded by the UK Particle Physics and Astronomy Research Council (PPARC), the JCMT Development Fund and the Canadian Foundation for Innovation (CFI).